\documentclass[manuscript,sigplan,screen]{acmart}
\usepackage{pgfplots}
\pgfplotsset{compat=1.18}
\usepackage{listings}
\usetikzlibrary{fit, positioning, arrows.meta, calc}
\AtBeginDocument{%
  }

\begin{document}

\title{Answer Set Programming for Egg Extraction and More}

\author{Ziyi Yang}
\email{yangziyi@u.nus.edu}
\affiliation{%
  \institution{National University of Singapore}
  \city{Singapore}
  \country{Singapore}
}

\author{Ilya Sergey}
\email{ilya@nus.edu.sg}
\affiliation{%
  \institution{National University of Singapore}
  \city{Singapore}
  \country{Singapore}
}


\newcommand{\todo}[1]{\textcolor{red}{[TODO: #1]}}

\begin{abstract}
Three years ago, Philip Zucker posted an attempt to use answer set programming (ASP) for term extraction from e-graphs~\cite{zucker2026answersetpro}. Although the task is NP-hard and ASP offers a natural modelling of e-graph terms, the initial attempt did not yield convincing results.

From the aspect of practical ASP users, we first pinpoint the way to make ASP work and work well on the task of e-graph extraction. The initial results show the na\"ive ASP encoding is comparable on efficiency to the well-optimised ILP-based exact DAG extraction in the extraction-gym, and find several extra optimal extraction on the complex instances. This leads us to a further agenda: with the ``better together of egg+Datalog'', is there a better ``better together'' by having ASP as a more powerful Datalog? We discuss the potential benefit from each other.
\end{abstract}

\maketitle

\section{DAG Extraction in E-Graphs}

Term extraction in e-graphs is a central optimisation step for equality saturation systems. The exact extraction problem is known to be NP-hard~\cite{egsat-nphard}, which makes solver choice and encoding strategy directly relevant for practical use. Several recent works have explored the solution from theory to practice~\cite{extraction-sparse,eboost,extraction-treewidth,extraction-sat}.

Prior work highlights that tree-cost intuition does not transfer directly to exact DAG extraction: sharing in DAGs creates global interactions that are not captured by purely local, compositional tree costs. Importantly, Philip's ASP formulation~\cite{zucker2026answersetpro} already targets DAG-cost extraction by using the stable-model semantics of ASP to enforce acyclicity.
That encoding is compact and expressive, but baseline experiments in the original post indicated practical scalability limitations on larger instances. This extended abstract revisits the same task with a solver-oriented ASP workflow.

\section{The Attempts and the Problems}

\subsection{How ASP Models E-Graph Extraction}
\begin{figure*}[t]
\centering
\begin{minipage}[t]{0.48\textwidth}
\textbf{Logical facts of an e-graph}
\vspace{2pt}
\begin{lstlisting}
% root eclass
root(1).

% enode(Eclass, Id, Op, Cost)
enode(1,0,"add",1).
enode(1,1,"add",1).
enode(2,0,"x",1).
enode(3,0,"y",1).

% child(Eclass, Id, ChildEclass)
child(1,0,2). child(1,0,3).
child(1,1,3). child(1,1,2).
\end{lstlisting}

\end{minipage}\hfill
\begin{minipage}[t]{0.45\textwidth}
\textbf{E-graph view (right)}
\vspace{2pt}
\centering
\begin{tikzpicture}[
    >=Stealth,
    eclass/.style={draw=blue!80, dashed, rounded corners, thick, inner sep=12pt},
    enode/.style={draw=black, solid, rectangle, rounded corners, thick, fill=gray!10, text centered, minimum height=2em, minimum width=3em},
    edge/.style={->, thick, draw=black!70},
    edgeLabel/.style={fill=white, inner sep=2pt, font=\scriptsize, text=black!80}
]

    \node[enode] (e2_0) {\texttt{"x"}};
    \node[eclass, fit=(e2_0)] (ec2) {};
    \node[above=8pt of ec2.east, text=blue!80, font=\small\bfseries] {E-class 2};
    \node[below=2pt of e2_0, font=\tiny, text=gray] {id: 0};

    \node[enode, right=4cm of e2_0] (e3_0) {\texttt{"y"}};
    \node[eclass, fit=(e3_0)] (ec3) {};
    \node[above=8pt of ec3.west, text=blue!80, font=\small\bfseries] {E-class 3};
    \node[below=2pt of e3_0, font=\tiny, text=gray] {id: 0};

    \path (ec2) -- (ec3) coordinate[midway] (mid);
    
    \node[enode, above left=2.5cm and 0.5cm of mid] (e1_0) {\texttt{"add"}};
    \node[enode, above right=2.5cm and 0.5cm of mid] (e1_1) {\texttt{"add"}};
    
    \node[eclass, fit=(e1_0)(e1_1)] (ec1) {};
    \node[above=2pt of ec1.north, text=blue!80, font=\small\bfseries] {E-class 1};
    
    \node[above=2pt of e1_0, font=\tiny, text=gray] {id: 0};
    \node[above=2pt of e1_1, font=\tiny, text=gray] {id: 1};

    
    \draw[edge] (e1_0.south) .. controls ++(down:1.5cm) and ++(up:1cm) .. (ec2.north) 
        node[edgeLabel, pos=0.5] {child 1};
    \draw[edge] (e1_0.south) .. controls ++(down:1.5cm) and ++(up:1.5cm) .. (ec3.north) 
        node[edgeLabel, pos=0.5] {child 2};

    \draw[edge] (e1_1.south) .. controls ++(down:1.5cm) and ++(up:1cm) .. (ec3.north) 
        node[edgeLabel, pos=0.5] {child 1};
    \draw[edge] (e1_1.south) .. controls ++(down:1.5cm) and ++(up:1.5cm) .. (ec2.north) 
        node[edgeLabel, pos=0.5] {child 2};

\end{tikzpicture}
\end{minipage}

\caption{A commutativity example ($x+y \equiv y+x$) shown as ASP facts and as an e-graph from \cite{zucker2026answersetpro}.}
\Description{Two side-by-side panels. Left panel lists ASP facts with root class 1, two add enodes in that class, and child links in opposite order, plus leaf classes for x and y. Right panel is a bipartite e-graph: class e1 points to two add enodes, classes e2 and e3 point to x and y leaves, and the two add enodes point to e2 and e3 in opposite argument order.}
\label{fig:facts-egraph}
\end{figure*}
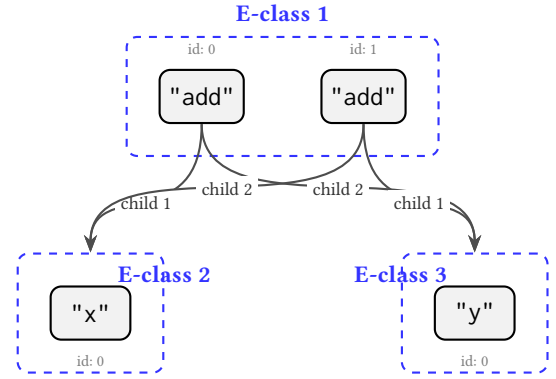

The key attraction of Philip's encoding is that it captures DAG extraction with a tiny set of relations and founded rules, without explicitly encoding loops. The e-graph is represented as facts in logic programming over classes and enodes, as shown in Figure~\ref{fig:facts-egraph}: each enode is identified inside an eclass (operator and local cost), and child links point from an enode to child eclasses. In practice this can be read as three tables: \texttt{root(E)}, \texttt{enode(E,I,Op,C)}, and \texttt{child(E,I,Ec)}.

The essential rule pattern is a support-style choice: an enode may be selected only if its child eclasses are selected; selecting an enode induces selection of its eclass; and every root eclass must be selected. Together with cost minimization, this yields a founded selected subgraph. The important point is that cycles are avoided through stable-model foundedness (no self-supporting selection), rather than by adding an explicit reachability or transitive-closure encoding. 

\begin{figure*}[t]
\centering
\begin{minipage}[t]{0.48\textwidth}
\textbf{Philip's bottom-up encoding}
\vspace{2pt}
\begin{lstlisting}[numbers=left,stepnumber=1,numbersep=6pt,firstnumber=1]
% choose an enode only if all child classes are selected
{sel(E,I)} :- enode(E,I,_,_), selclass(Ec): child(E,I,Ec).

% selecting an enode selects its class
selclass(E) :- sel(E,_).

% at most one enode per class
:- enode(E,_,_,_), #count { I : sel(E,I) } > 1.

% optimize extraction cost
#minimize { C,E,I : sel(E,I), enode(E,I,_,C) }.

% roots must be selected
:- root(E), not selclass(E).
\end{lstlisting}
\end{minipage}\hfill
\begin{minipage}[t]{0.48\textwidth}
\textbf{A top-down alternative?}
\vspace{2pt}
\begin{lstlisting}[numbers=left,stepnumber=1,numbersep=6pt,firstnumber=1]
% Top-down: root classes need extraction
need(E) :- root(E).

% For each needed class, choose exactly one node
{ selnode(I) : enode(E,I,_,_) } = 1 :- need(E).

% Selecting a node makes its children's classes needed too
need(E) :- selnode(I), echild(I,E).

% The actual cost function
#minimize { C,E,I : selnode(I), enode(E,I,_,C) }.
\end{lstlisting}
\end{minipage}
\caption{ASP encoding of egg extraction (left: bottom-up from \cite{zucker2026answersetpro}, right: top-down ignoring DAG acyclicity).}
\Description{Two-column code figure. Left column contains the support-conditioned choice rule, the rule that selected enodes imply selected classes, and the root-selection integrity constraint. Right column contains a one-enode-per-class integrity constraint and the minimization objective over selected enode costs.}
\label{fig:core-encoding}
\end{figure*}

Figure~\ref{fig:core-encoding} (left) shows the core rules for the initial bottom-up encoding from \cite{zucker2026answersetpro}.
Line 2 in the left column enforces support before selecting an enode, line 5 lifts selected enodes to selected eclasses, and line 14 makes root selection mandatory. Line 8 adds a one-enode-per-class restriction, while line 11 sets the optimisation objective.

\paragraph{The problem} It is unfortunate that the initial encoding ``appears to be slow on the large problems'' (as the original post states). For example, the \texttt{lambda function repeat} instance takes forever to solve in the ASP encoding, while even the na\"ive ILP encoding can solve it in less than a second. 



\subsection{A Top-Down Alternative?}
Although not discussed explicitly, the natural question is whether the top-down alternative for the extraction is valid. Figure~\ref{fig:core-encoding} (right) sketches a top-down encoding: root classes are marked as needed (line 2), and for each needed class, exactly one enode is selected (line 5). Selecting an enode makes its children needed too (line 8), which creates a demand-driven support condition. The same cost minimization applies to the selected nodes (line 11).

\paragraph{The problem} Unlike the bottom-up encoding, the search from the roots fails to skip the cycles in the e-graph, which means incorrectness of the DAG extraction. The question is whether we can fix this easily with ASP constraints.

A natural attempt is to add a reachability constraint: if an enode is selected, we collect all its reachable nodes from the selected enodes, and require that none of the selected nodes are reachable from themselves. A direct encoding is like

\begin{small}
\begin{verbatim}
% Selected dependency graph
sel_edge(E,Ec) :- selnode(I), enode(E,I,_), echild(I,Ec).

% Transitive closure of selected dependency graph
sel_reach(E,Ec) :- sel_edge(E,Ec).
sel_reach(E,Ez) :- sel_reach(E,Ey), sel_edge(Ey,Ez).

% Cycle prohibition: extracted structure must be acyclic
:- sel_reach(E,E).
\end{verbatim}
\end{small}

However, this encoding is not effective since the real e-graphs can have large strongly connected components (SCCs) with many nodes, making the $O(n^3)$ transitive closure encoding too heavy for the solver.

\section{The Pragmatic Solutions}

We say it is unfortunate (about the poor scalability), because Philip was almost there: if he had tried using parallel solving~\cite{clingo-multithreaded}, even with two threads, ASP version should work. This is actually unfortunate due to the barrier of Clingo~\cite{clingo-modern}, the most popular ASP solver. What is helping is actually the \emph{UNSAT-core} (\emph{a.k.a} usc) based optimisation in Clingo~\cite{clingo-usc}, which is a powerful technique for solving optimisation problems (and enabled by the second thread with parallel solving). With single-thread solving on usc, the ASP encoding three years ago is competitive with the latest ILP-based exact DAG extraction, for which we will show the results later.

The solution for the top-down encoding is more involved in principle: to detect and eliminate cycles is not complex, but the declarative  logic programming is limited in expressing such procedures. The good thing is, Clingo provides custom propagators~\cite{clingo-modern}. So like how SMT solvers use theory solvers, the propagator can be used to implement imperative procedures to support the solving, which is answer set programming modulo acyclicity in this case. The funny thing is, such propagator exists in Clingo (but without documentation). So it only takes one more line to achieve the SCC elimination:
\begin{small}
\begin{verbatim}
#edge (E,Ec): selnode(I), enode(E,I,_), echild(I,Ec).
\end{verbatim}
\end{small}

It is exactly the same edge definition as the one above, but the reachability is handled outside the ASP solver.

\section{Benchmarking}

We evaluate against extraction-gym\footnote{https://github.com/egraphs-good/extraction-gym} baselines on three suites (babble, herbie, rover), which are a large collection of e-graphs from different domains. We use \emph{faster-greedy-dag} as the baseline for solution quality, and compare the top-down and bottom-up ASP encodings against the na\"ive ILP encoding and an optimised ILP variant (with several optimisations outside the ILP solving). All exact methods are given a default 10-second timeout in the extraction-gym.

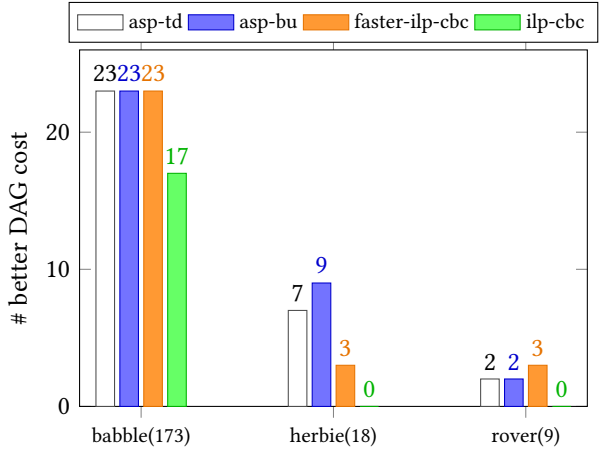
\begin{figure}[!t]
\centering
\begin{tikzpicture}
\begin{axis}[
  ybar,
  bar width=7pt,
  area legend,
  width=0.98\linewidth,
  height=6.3cm,
  enlarge x limits=0.16,
  ymin=0,
  ymax=26,
  ylabel={\# better DAG cost},
  symbolic x coords={babble,herbie,rover},
  xtick=data,
  xticklabels={babble(173),herbie(18),rover(9)},
  xticklabel style={align=center,font=\footnotesize},
  legend style={font=\footnotesize, at={(0.5,1.02)},anchor=south,legend columns=4},
  nodes near coords,
  nodes near coords align={vertical},
]
\addplot+[fill=white, draw=black!70, nodes near coords style={text=black}] coordinates {(babble,23) (herbie,7) (rover,2)};
\addplot+[fill=blue!55, draw=blue!80!black, nodes near coords style={text=blue!80!black}] coordinates {(babble,23) (herbie,9) (rover,2)};
\addplot+[fill=orange!80, draw=orange!90!black, nodes near coords style={text=orange!90!black}] coordinates {(babble,23) (herbie,3) (rover,3)};
\addplot+[fill=green!60, draw=green!70!black, nodes near coords style={text=green!70!black}] coordinates {(babble,17) (herbie,0) (rover,0)};
\legend{asp-td,asp-bu,faster-ilp-cbc,ilp-cbc}
\end{axis}
\end{tikzpicture}
\caption{Per-suite counts of instances with strictly better DAG cost than \texttt{faster-greedy-dag}. Suite sizes are shown on the x-axis.}
\Description{Grouped bar chart over babble, herbie, and rover showing how many instances each method improves over faster-greedy-dag. The top-down ASP bar is white; the bottom-up ASP bar is blue; faster-ilp-cbc-timeout is orange; ilp-cbc-timeout is green. Counts are 23, 7, and 2 for top-down ASP, 23, 9, and 2 for bottom-up ASP, 23, 3, and 3 for faster-ilp-cbc-timeout, and 17, 0, and 0 for ilp-cbc-timeout.}
\label{fig:dag-better}
\end{figure}

\begin{figure}[!t]
\centering
\begin{tikzpicture}
\begin{axis}[
  ybar,
  bar width=12pt,
  area legend,
  width=0.98\linewidth,
  height=6.3cm,
  enlarge x limits=0.16,
  ymin=0,
  ymax=3.6,
  ylabel={geo-mean runtime ratio},
  symbolic x coords={babble,herbie,rover},
  xtick=data,
  legend style={font=\footnotesize, at={(0.5,1.02)},anchor=south,legend columns=3},
  nodes near coords,
  nodes near coords align={vertical},
]
\addplot+[fill=blue!55, draw=blue!80!black, nodes near coords style={text=blue!80!black}] coordinates {(babble,1.5074) (herbie,0.9569) (rover,1.1245)};
\addplot+[fill=orange!80, draw=orange!90!black, nodes near coords style={text=orange!90!black}] coordinates {(babble,0.7797) (herbie,1.9442) (rover,0.4924)};
\addplot+[fill=green!60, draw=green!70!black, nodes near coords style={text=green!70!black}] coordinates {(babble,3.3231) (herbie,3.2567) (rover,1.6715)};
\legend{asp-bu / asp-td,faster-ilp-cbc / asp-td,ilp-cbc / asp-td}
\end{axis}
\end{tikzpicture}
\caption{Geometric-mean runtime ratio by suite from pair comparisons to the top-down \texttt{asp-td} encoding. Values below 1 indicate faster than \texttt{asp-td}.}
\Description{Grouped bar chart of geometric-mean runtime ratios versus asp-td across babble, herbie, and rover. Bottom-up ASP ratios are 1.5074, 0.9569, and 1.1245. faster-ilp-cbc-timeout ratios are 0.7797, 1.9442, and 0.4924. ilp-cbc-timeout ratios are 3.3231, 3.2567, and 1.6715. Lower values are better.}
\label{fig:runtime-ratio}
\end{figure}
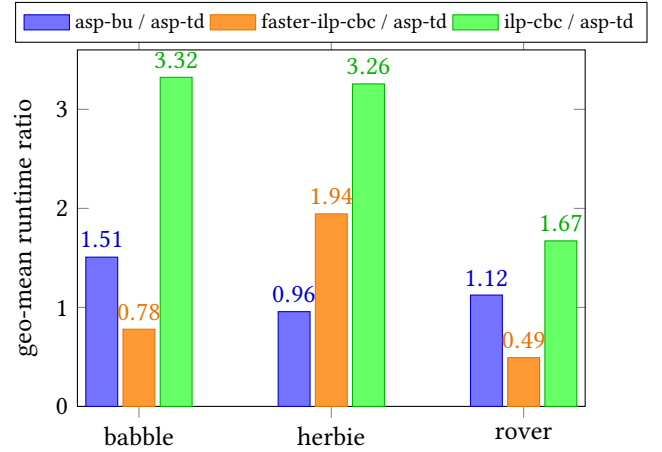

Figure~\ref{fig:dag-better} summarizes solution quality as the number of instances where each method improves DAG cost over the greedy. Runtime trade-offs are shown in Figure~\ref{fig:runtime-ratio} as geometric-mean ratios against the \texttt{asp-td} (lower is better). The top-down series is omitted from the plot, while the bottom-up ASP and the ILP variants are shown for comparison.

Overall, the propagator-based top-down encoding (asp-td) gives a good balance in our evaluation: it preserves solution quality while staying competitive in runtime. By contrast, the na\"ive ILP baseline is clearly not competitive.

The optimised ILP variant can be very fast, especially on rover. Our interpretation is that this comes largely from non-solver engineering choices (including specialized optimisations such as zero-cost-oriented simplifications) that can significantly help certain classes of instances. With Clingo's API, similar engineering hooks are also feasible for ASP.

Compared with asp-bu, asp-td is generally faster overall, but the quality differences are concentrated in three outlier instances, two of which are additional optimal extractions found by asp-bu. These outliers are root-heavy (around 400 and 1400 roots), where bottom-up search can be advantageous because it does not start from roots. At the same time, the runtime benefit of asp-bu on the suites where it helps remains modest (for example, geo-mean ratio is only about 0.96 on herbie), so asp-bu is still slower overall. This indicates that search direction is strongly case-sensitive.

There are several limitations to the current evaluation. First, the weak constraints for optimisation in Clingo only cover integer costs, which means that the current ASP encodings are not directly applicable to non-integer cost e-graphs. Second, we did not investigate more solver configurations and features: the only extra one we tried is a 4-thread parallel solving for Clingo, which only achieves better suboptimal results but not better optimal results in our experiments. A portfolio-based approach may be a good way to leverage different encodings, optimisation techniques, and solver features for different instance classes.

\section{More than Extraction?}

The immediate takeaway is practical: the two-phase egg extraction--first egg then extraction--is effective and efficient with the right ASP encoding and solver configuration.
Looking forward, this opens a broader agenda around solver-enhanced equality saturation: if egg and Datalog are ``better together,'' then ASP may provide an even richer integration point by combining declarative modelling with stronger global optimisation and constraint handling. In the following, we discuss our ongoing work on this agenda.

\subsection{What does ASP do?}

It searches--in a generate-and-test style--searching for
$$\exists M.\ M\in\mathit{G} \land \mathit{T}(M)$$
where $\mathit{G}$ is the search space defined by the ASP encoding, and
$\mathit{T}$ is the test condition defined by the constraints (more
layers of quantification in the case of optimisation problems).

\subsection{What does EqSat do?}

It compactly represents a congruence-closed search space of equivalent
terms. An e-graph maintains a set of e-classes, each partitioning a
collection of e-nodes (operator applications) that are known to be
equivalent under the current rewrite state.
The central congruence invariant is: \emph{equivalent children induce
equivalent parents, so every term extractable from an e-class is
equivalent to every other term extractable from the same e-class.}
Rewrite rules extend the e-graph by merging e-classes, growing the
space of represented terms while preserving this invariant throughout.


\subsection{The Combined Formulation of Extraction}

This makes the two frameworks naturally complementary. In egglog~\cite{egglog}, the combination
provides a compact, equivalence-closed search space
$$G_{\mathit{egraph}} = \mathit{lfp}(\mathit{Datalog}_E)(\mathit{facts}(t_0))$$
computed as the Datalog least fixpoint of the rewrite rules, while ASP
navigates that space with full combinatorial search and optimisation:
$$\exists M \in G_{\mathit{egraph}}.\; T(M) \quad
  [\,\text{minimise}\ \mathit{cost}(M)\,]$$

The gap here is clear: in ASP, the search space is defined by non-deterministic rules rather than the definite rules enabling monotone evaluation in Datalog, so the congruence closure is to be achieved differently, not as naturally as in Datalog.
ASP's role is then not
merely extraction after the fact, but potentially a richer integration:
its CDCL engine can prune saturation by short-circuiting rule firing
when a sufficiently good $M$ is already reachable, and its weak
constraints can guide \emph{which} rewrites are worth exploring.
This points toward a single, cost-aware saturation-plus-extraction
pipeline rather than a strict two-phase decomposition.

\subsection{Equivalence Checking as Incremental Search}

Equivalence checking admits a different instantiation that falls
\emph{outside} the static generate-and-test quantification above.
Rather than searching over a fixed space $G_{\mathit{egraph}}$,
checking asks whether $s$ and $t$ can be merged by \emph{some} finite
sequence of rule firings---a question whose answer depends on the
evolving state of the e-graph itself.

This maps naturally to incremental solving: each step extends the
e-graph by one bounded increment of rule applications and checks
whether $[s] = [t]$ holds in the current state. Crucially, each
intermediate state $G_k$ is not a raw accumulation of derived terms
but is \emph{compressed} by congruence closure---equivalent terms are
merged and represented once, so the search horizon grows in the space
of equivalence classes rather than individual terms. Formally, let
$G_k$ denote the congruence-closed e-graph after $k$ increments; the
checking problem becomes:
$$\exists k.\; [s]_{G_k} = [t]_{G_k}$$
with termination as soon as the condition is met---or when $G_k =
G_{k-1}$ (fixpoint reached without witnessing the merge, certifying
$s \not\equiv_E t$). 

This is structurally analogous to bounded model checking: each
increment deepens the rewriting horizon by one step, and the
equivalence  plays the role of the safety property.

\subsection{How EqSat might enhance ASP?}


In the scenarios above, full equality saturation may be unnecessary—or infeasible within resource bounds. Instead, CDCL-based ASP systems can evolve the search space on the fly, firing rewrites lazily only when they are needed to improve or certify a candidate solution. The more ambitious reverse direction is to use an e-graph to enhance ASP: an ``ASP modulo equivalence'' workflow where congruence closure provides additional propagation and pruning during stable-model search. How to expose and exploit such congruence information effectively inside current ASP solving remains an open question.

\bibliographystyle{ACM-Reference-Format}
\bibliography{sample-base}

\end{document}